\documentclass{article}
\usepackage{amsmath}

\input{tcilatex}

\begin{document}

\title{Gravitinos Tunneling from Black Holes}
\author{Alexandre Yale\thanks{%
ayale@uwaterloo.ca} ~and~R. B. Mann\thanks{%
rbmann@sciborg.uwaterloo.ca} \\
%EndAName
Department of Physics \& Astronomy, University of Waterloo\\
Waterloo, Ontario N2L 3G1, Canada}
\date{August 13, 2008}
\maketitle

\abstract{Black hole radiation of gravitinos is investigated as the
classically forbidden tunneling of spin-3/2 fermions through an event
horizon. We show that all four spin states of the gravitino yield the same
emission temperature, and retrieve the Unruh temperature in a Rindler
spacetime as well as the Hawking temperature for a Kerr-Newman charged
rotating black hole.}

%%%%%%%%%%%%%%%%%%%%%%%%%%%%%%%%%%%%  INTRODUCTION  %%%%%%%%%%%%%%%%%%%%%%%%%%%%%%%%%%%%%%%%%%%%%%%%%%%%%%%%

\section{Introduction}

In recent years, theorists have begun modeling Hawking radiation as a
tunneling effect, offering an alternate conceptual understanding to black
hole radiation. Originating with the early work of Kraus and Wilczek \cite%
{early tunnelling}, the tunneling approach was later refined \cite{late 90s
tunnelling,Parikh, Padmanabhan} to show that black hole radiation can indeed
be understood in terms of local physics near the horizon. This approach is
quite distinct from more global methods such as Wick Rotation methods and
Hawking's original method of modeling gravitational collapse \cite{hawking}.
Although the approach was first demonstrated for scalar particles tunneling
out of spherically symmetric Schwarzschild black holes, it has been shown to
be remarkably versatile, and has been applied successfully to a wide variety
of spacetimes, including Kerr and Kerr-Newmann cases \cite{kerr and kerr
newman, Zhang and Zhao, first paper}, black rings \cite{Black Rings}, the
3-dimensional BTZ black hole \cite{Vanzo, BTZ}, Vaidya \cite{Vaidya}, other
dynamical black holes \cite{dynamicalbh}, Taub-NUT spacetimes \cite{first
paper}, and G\"{o}del spacetimes \cite{Godel}. One can even recover the
Unruh temperature \cite{Unruh} for a Rindler observer \cite%
{Padmanabhan,first paper} using this approach.

All tunneling approaches make use of the WKB approximation to compute the
tunneling probability for a classically forbidden trajectory across the
horizon. In the present paper, we will consider an infinitesimal path
directed in the radial direction from the inside to the outside of the
horizon. For the semiclassical action $I$ of this trajectory to leading
order in $\hbar $ (which we set to unity), this yields \cite{early
tunnelling,Massar}%
\begin{equation}
\Gamma \propto \exp (-2\,\mathrm{Im}I)  \label{Tunprob}
\end{equation}%
One expects that a black hole should radiate all types of particles, akin to
a black body at a well-defined temperature (ignoring grey body effects),
such that particles of every spin should appear in the emission spectrum.
While the implications of this were studied 30 years ago \cite{don page}, it
has only recently been shown that spin-1/2 fermions can tunnel out of black
holes \cite{first paper}, yielding (to leading order in a WKB approximation)
the same temperature as one obtains for scalar radiation. As with the scalar
case, this approach has also been shown to be quite robust, and has been
applied to a broad variety of black hole spacetimes \cite{Fermion
tunnelling,ferm BTZ,ferm variable mass}.

In this paper we consider radiation of spin-3/2 fermions, or gravitinos.
Such particles are predicted to exist in all theories of supergravity \cite%
{sugraref}. Since local supersymmetry is not a symmetry of the state of our
observable universe, it should be broken, and so the gravitino should be
massive. As the gravitino obeys the Rarita-Schwinger Equation \cite%
{PhysRev.60.61}, any one of four possible spin states can be emitted by the
black hole. With reference to the radial direction, these components are
3/2, 1/2, -1/2, and -3/2. Since it is not a-priori clear that all components
are emitted in the same way, we have a new and interesting check on the
tunneling method.

Our approach will be analogous to that taken in the spin-1/2 case \cite%
{Fermion tunnelling,kerner-2008}. Employing a WKB approximation to the
Rarita-Schwinger Equation, we will show that a general structure emerges
that ensures all four spin states are radiated by the black hole with at the
same Hawking temperature. To keep the discussion more general, we will
couple the gravitino to a U(1) gauge field as well as gravity. We apply
these results first to Rindler spacetime, confirming that the Unruh
temperature is recovered. We then consider the Kerr-Newman case, which had
introduced some non-trivial technical features associated with the choice of 
$\gamma $ matrices for the spin-1/2 case \cite{kerner-2008}. We show that
the same approach is sufficient in describing gravitino emission to leading
order in the WKB approximation\footnote{%
Our approach to gravitino radiation differs from that taken by Jing, who
used the Newman-Penrose formalism for a Reissner-Nord\-str\"{o}m metric \cite%
{Jing} to show that the radiating gravitinos necessarily obey a 2nd-order
Klein-Gordon-type differential equation in this background}.

One of the assumptions of our semi-classical calculation is to neglect any
change of angular momentum of the black hole due to the spin of the emitted
particle. \ For zero-angular momentum black holes with mass much larger than
the Planck mass this is a good approximation. Furthermore, particles of
opposite spin will be emitted in equal numbers statistically, yielding no
net change in the angular momentum of the black hole (although second-order
statistical fluctuations will be present). We confirm that spin 3/2 fermions
are also emitted at the Hawking Temperature. This final result, while not
surprising, furnishes an important confirmation of the robustness of the
tunneling approach. \ 

\section{Gravitinos in a Black Hole Background}

The Rarita-Schwinger equation representing the spin-$3/2$ fermion field will
be used in the form \cite{PhysRev.60.61} 
\begin{subequations}
\begin{gather}
i\gamma ^{\nu }\left( D_{\nu }+iqA_{\nu }\right) \Psi _{\mu }+m\Psi _{\mu }=0
\label{eqn:RS2} \\
\gamma ^{\mu }\Psi _{\mu }=0,  \label{eqn:RS2b}
\end{gather}%
where $\Psi _{\mu }\equiv \Psi _{\mu a}$ is a vector-valued spinor of charge 
$q$ and mass $m$, $A_{\nu }$ represents the electromagnetic potential of the
black hole, and the $\gamma ^{\mu }$ matrices satisfy $\left\{ \gamma ^{\mu
},\gamma ^{\nu }\right\} =2g^{\mu \nu }$. The covariant derivative obeys 
\end{subequations}
\begin{equation}
\begin{gathered} D_\mu = \partial_\mu + \Omega_\mu \\ \Omega_\mu =
\frac{1}{2} i \Gamma^{\alpha \beta}_\mu \Sigma_{\alpha \beta}\\
\Sigma_{\alpha \beta} = \frac{1}{4} i [\gamma^\alpha,\gamma^\beta],
\end{gathered}
\end{equation}%
where $\Omega _{\mu }$ is the spin-connection.

The first equation is the Dirac equation applied to every vector index of $%
\Psi$, while the second is a set of additional constraints to ensure that no
ghost state propagates; that is, to ensure that $\Psi$ represents only
spin-3/2 fermions, with no spin-1/2 mixed states.

Working in the context of the path-integral formalism and the WKB
approximation, it is known that each path will have a phase of $\exp
(iI/\hbar )$, where $I$ is the action corresponding to that path. In the
case that we consider here, the infinitesimal radial path across the horizon
will dominate, such that we may employ the following ansatz for the wave
function: 
\begin{equation}
\Psi _{\mu }=\left[ 
\begin{array}{c}
a_{\mu } \\ 
b_{\mu } \\ 
c_{\mu } \\ 
d_{\mu }%
\end{array}%
\right] e^{\frac{i}{\hbar }I},  \label{ansatz}
\end{equation}%
where the $a_{\mu },b_{\mu },c_{\mu },d_{\mu }$ are each functions of the
spacetime.

Note that the first set of Rarita-Schwinger equations (\ref{eqn:RS2}) will
yield an equation which can be solved for the action $I$ independently of
the wave function components $a_{\mu },\ldots ,d_{\mu }$. The second set of
Rarita-Schwinger equations (\ref{eqn:RS2b}), on the other hand, will yield
four constraints for these wave function components independently of the
action, and will have solutions in every spacetime. Hence, as the action is
all that we require to find the emission temperature, we need solve only
equation (\ref{eqn:RS2}). This implies that fermions of every spin will emit
at the same temperature, since equation (\ref{eqn:RS2b}) will never have any
effect on the action $I$.

Defining the $\hat{\gamma}$ matrices to be of the chiral form, 
\begin{equation}
\hat{\gamma}^{0}= \left[ 
\begin{array}{cc}
0 & 1_{2} \\ 
-1_{2} & 0%
\end{array}%
\right] \hspace{1cm}\hat{\gamma}^{i}= \left[ 
\begin{array}{cc}
0 & \sigma ^{i} \\ 
\sigma ^{i} & 0%
\end{array}%
\right] ,  \label{eqn:gamma}
\end{equation}%
where $\gamma ^{\mu }=e_{I}^{\mu }\hat{\gamma}^{I}$ are metric-dependent
linear combinations of these matrices, we find that the first
Rarita-Schwinger equations (\ref{eqn:RS2}) can be rewritten in the form 
\begin{equation}
\left[ 
\begin{array}{cccc}
m &  & x_{0}+x_{3} & x_{1}-ix_{2} \\ 
& m & x_{1}+ix_{2} & x_{0}-x_{3} \\ 
-x_{0}+x_{3} & x_{1}-ix_{2} & m &  \\ 
x_{1}+ix_{2} & -x_{0}-x_{3} &  & m%
\end{array}%
\right] \left[ 
\begin{array}{c}
a \\ 
b \\ 
c \\ 
d%
\end{array}%
\right] =0,
\end{equation}%
where 
\begin{equation}
x_{a}=e_{a}^{\nu }\partial _{\nu }I.
\end{equation}

Performing an LU-Decomposition of the matrix above yields a solution
independent of the wave function components: 
\begin{equation}
\eta ^{ab}\left( e_{a}^{\mu }\partial _{\mu }I\right) \left( e_{b}^{\mu
}\partial _{\mu }I\right) -m^{2}=0,  \label{eqn:fermionaction}
\end{equation}%
where $\eta =\text{diag}(-1,1,1,1)$ is the Minkowski flat-space metric. This
equation is equivalent to the first set of Rarita-Schwinger equations (\ref%
{eqn:RS2}), and will allow us to solve for $\partial _{z}I$, regardless of
what $\Psi $ looks like, and hence independently of (\ref{eqn:RS2b}).
Integrating $\partial _{z}I$ along the infinitesimal radial path across the
event horizon will yield a complex residue corresponding to the action of
the radiation. From this action, the temperature is found using \cite{early
tunnelling,Massar} 
\begin{equation}
\Gamma \propto \exp \left( -2 \text{Im} I\right) \propto \exp \left( \frac{-E%
}{T_{H}}\right) .
\end{equation}%
Note that the above relations hold regardless of the spin-state chosen for
gravitino emission. Consequently, all four spin states are radiated with the
same Hawking temperature to leading order in WKB. We will now explicitly
demonstrate this method in Rindler and Kerr-Newman spacetimes.

\section{Rindler Spacetime}

We first make use of the results of the preceding section in Rindler
spacetime, whose simplicity allows one to easily grasp the concept of the
method employed. We will here show that it allows us to retrieve the Unruh
Temperature \cite{Unruh} for gravitinos. Consider the metric 
\begin{equation}
ds^{2}=-f(z)dt^{2}+dx^{2}+dy^{2}+\frac{dz^{2}}{g(z)}  \label{Rindmet}
\end{equation}%
where 
\begin{equation}
\begin{split}
f(z)& =\alpha ^{2}z^{2}-1 \\
g(z)& =\frac{\alpha ^{2}z^{2}-1}{\alpha ^{2}z^{2}} \\
A_{\mu }& =0,
\end{split}%
\end{equation}%
which has a singularity at $z=\alpha ^{-1}$. We will use the $\gamma ^{\mu }$
matrices in the form 
\begin{equation}
\gamma ^{t}=\frac{1}{\sqrt{f(z)}}\hat{\gamma}^{0}\hspace{0.5cm}\gamma ^{x}=%
\hat{\gamma}^{1}\hspace{0.5cm}\gamma ^{y}=\hat{\gamma}^{2}\hspace{0.5cm}%
\gamma ^{z}=\sqrt{g(z)}\hat{\gamma}^{3},
\end{equation}%
and consider the action to have a solution to (\ref{eqn:fermionaction}) of
the form 
\begin{equation}
I=-Et+W(z)+P(x,y)+K.  \label{eqn:rindleraction}
\end{equation}%
Inserting (\ref{eqn:rindleraction}) into the Rarita-Schwinger equation (\ref%
{eqn:RS2}), we obtain, to leading order in $\hbar $ such that neither $%
\Omega _{\mu }$ nor the derivatives of the $a_{\mu },\ldots ,d_{\mu }$
contribute, the four equations: 
\begin{subequations}
\label{eqn:Rindler}
\begin{gather}
c_{\mu }\left( \frac{E}{\sqrt{f}}-W^{\prime }\sqrt{g}\right) -d_{\mu }\left(
P_{x}-iP_{y}\right) +a_{\mu }m=0 \\
d_{\mu }\left( \frac{E}{\sqrt{f}}+W^{\prime }\sqrt{g}\right) -c_{\mu }\left(
P_{x}+iP_{y}\right) +b_{\mu }m=0 \\
a_{\mu }\left( \frac{E}{\sqrt{f}}+W^{\prime }\sqrt{g}\right) +b_{\mu }\left(
P_{x}-iP_{y}\right) -c_{\mu }m=0 \\
b_{\mu }\left( \frac{E}{\sqrt{f}}-W^{\prime }\sqrt{g}\right) +a_{\mu }\left(
P_{x}+iP_{y}\right) -d_{\mu }m=0.
\end{gather}%
The constraints (\ref{eqn:RS2b}) will also give us the relations 
\end{subequations}
\begin{subequations}
\label{eqn:Rindler2}
\begin{gather}
\frac{c_{1}}{\sqrt{f}}+d_{2}-id_{3}+\sqrt{g}c_{4}=0 \\
\frac{d_{1}}{\sqrt{f}}+c_{2}+ic_{3}-\sqrt{g}d_{4}=0 \\
\frac{-a_{1}}{\sqrt{f}}+b_{2}-ib_{3}+\sqrt{g}a_{4}=0 \\
\frac{-b_{1}}{\sqrt{f}}+a_{2}+ia_{3}-\sqrt{g}b_{4}=0
\end{gather}%
between the various components of the wave function. These are not important
here as our equations will yield a solution for the action that is
independent of the wave function, meaning that these new constraints cannot
have any effect on the action provided they allow a non-zero solution $\Psi $%
, which they always will.

Solving equations (\ref{eqn:Rindler}) exactly and generally, we get the same
solution as (\ref{eqn:fermionaction}): 
\end{subequations}
\begin{equation}
\left( \frac{E}{\sqrt{f}}-W^{\prime }\sqrt{g}\right) \left( \frac{E}{\sqrt{f}%
}+W^{\prime }\sqrt{g}\right) -(P_{x}-iP_{y})(P_{x}+iP_{y})+m^{2}=0.
\end{equation}%
As we approach the horizon, we find that $f(z)$ and $g(z)$ both go to zero,
while the other terms (besides $W^{\prime }$) are constant, and therefore
get 
\begin{equation}
W_{\pm }^{\prime }=\frac{\pm E}{\sqrt{fg}}.  \label{eqn:RindlerSolution}
\end{equation}%
This solution will be valid for any non-zero $\Psi $ satisfying (\ref%
{eqn:Rindler2}), whose solution space generally is 12-dimensional.

We then integrate (\ref{eqn:RindlerSolution}) over our path through the
event horizon, which contains a pole that will be the sole imaginary
contribution to the action. This yields: 
\begin{equation}
ImW_{\pm }(z)=\frac{\pm \pi E}{\sqrt{f_{z}(z_{0})g_{z}(z_{0})}}=\frac{\pi E}{%
2\alpha }.
\end{equation}%
As $W^{\prime }$ corresponds to the momentum of the particle, one finds that 
$W_{+}$ corresponds to an outgoing particle, whereas $W_{-}$ corresponds to
an incoming one. By forcing to unity the probability that the incoming
particle is absorbed and using the fact that $W_{+}=-W_{-}$, we find 

\begin{equation}
\begin{split}
\Gamma \propto \exp (-2ImI)& =\frac{\exp \left(
-2(ImW_{+}+ImP(x,y)+ImK)\right) }{\exp \left(
-2(ImW_{-}+ImP(x,y)+ImK)\right) } \\
& =\exp (-4ImW_{+}) \\
& =\exp \left( \frac{-2\pi }{\alpha }E\right) ,
\end{split}%
\end{equation}%
which gives us the expected Unruh temperature \cite{Unruh}: 
\begin{equation}
T_{H}=\frac{\alpha }{2\pi }.
\end{equation}

\section{Kerr-Newman Black Hole Spacetime}

We next consider applying this method to a general Kerr-Newman black hole
spacetime. The metric is given by 
\begin{equation}
ds^{2}=-fdt^{2}+\frac{dr^{2}}{g}-2Hdtd\phi +Kd\phi ^{2}+\Sigma d\theta ^{2},
\end{equation}%
where 
\begin{equation}
\begin{split}
A_{\mu }& =\frac{-er}{\Sigma (r)}\left( (dt)_{\mu }-\alpha sin^{2}(\theta
)(d\phi )_{\mu }\right)  \\
f(r,\theta )& =\frac{\Delta (r)-\alpha ^{2}sin^{2}(\theta )}{\Sigma
(r,\theta )} \\
g(r,\theta )& =\frac{\Delta (r)}{\Sigma (r,\theta )} \\
H(r,\theta )& =\frac{\alpha sin^{2}(\theta )(r^{2}+\alpha ^{2}-\Delta (r))}{%
\Sigma (r,\theta )} \\
K(r,\theta )& =\frac{(r^{2}+\alpha ^{2})^{2}-\Delta (r)\alpha
^{2}sin^{2}(\theta )}{\Sigma (r,\theta )}sin^{2}(\theta ) \\
\Sigma (r,\theta )& =r^{2}+\alpha ^{2}cos^{2}(\theta ) \\
\Delta (r)& =r^{2}+\alpha ^{2}+e^{2}-2Mr=(r-r_{-})(r-r_{+}).
\end{split}%
\end{equation}%
We assume a non-extremal black hole, $M^{2}>\alpha ^{2}+e^{2}$, such that we
have two horizons at 
\begin{equation}
r_{\pm }=M\pm \sqrt{M^{2}-\alpha ^{2}-e^{2}}.
\end{equation}%
To simplify the notation, we will also use the functions 
\begin{subequations}
\begin{gather}
F(r,\theta )=f(r,\theta )+\frac{H^{2}(r,\theta )}{K(r,\theta )}=\frac{\Delta
(r)\Sigma (r,\theta )}{(r^{2}+\alpha ^{2})^{2}-\Delta (r)\alpha
^{2}sin^{2}(\theta )} \\
\Omega _{H}=\frac{H(r_{+},\theta )}{K(r_{+},\theta )}=\frac{a}{%
r_{+}^{2}+\alpha ^{2}},
\end{gather}%
where $\Omega _{H}$ corresponds to the angular velocity of the black hole.
We will use the following $\gamma ^{\mu }$ matrices \cite{kerner-2008} : 
\end{subequations}
\begin{equation}
\begin{gathered} \gamma^t = \frac{1}{\sqrt{F(r,\theta)}} \hat{\gamma}^0
\hspace{1.cm} \gamma^r = \sqrt{g(r,\theta)} \hat{\gamma}^3 \hspace{1.cm}
\gamma^\theta = \frac{1}{\sqrt{\Sigma(r,\theta)}} \hat{\gamma}^1\\
\gamma^\phi = \frac{1}{\sqrt{K(r,\theta)}} \left( \hat{\gamma}^2 +
\frac{H(r,\theta)}{\sqrt{F(r,\theta)K(r,\theta)}} \hat{\gamma}^0 \right) ,
\end{gathered}
\end{equation}%
where the $\hat{\gamma}^{i}$ are the chiral Minkowski matrices (\ref%
{eqn:gamma}). The action will in this case take the form 
\begin{equation}
I=-Et+W(r,\theta )+J\phi .
\end{equation}%
If we then input our wave function (\ref{ansatz}) into the Rarita-Schwinger
equations (\ref{eqn:RS2}), we get, again to leading order in $\hbar $, such
that neither $\Omega _{\mu }$ nor the derivatives of the $a_{\mu },\ldots
,d_{\mu }$ contribute, equations similar to those found for the Rindler
spacetime: 
\begin{subequations}
\label{eqn:BH}
\begin{gather}
c\left( \zeta -W_{r}\sqrt{g}\right) +d\left( \xi -\frac{W_{\theta }}{\sqrt{%
\Sigma }}\right) +am=0 \\
d\left( \zeta +W_{r}\sqrt{g}\right) +c\left( -\xi -\frac{W_{\theta }}{\sqrt{%
\Sigma }}\right) +bm=0 \\
a\left( \zeta +W_{r}\sqrt{g}\right) -b\left( -\xi -\frac{W_{\theta }}{\sqrt{%
\Sigma }}\right) -cm=0 \\
b\left( \zeta -W_{r}\sqrt{g}\right) -a\left( \xi -\frac{W_{\theta }}{\sqrt{%
\Sigma }}\right) -dm=0,
\end{gather}%
where 
\end{subequations}
\begin{subequations}
\begin{gather}
\zeta \equiv \frac{E}{\sqrt{F}}+\frac{qer}{\Sigma \sqrt{F}}-\left( J+\frac{%
qer}{\Sigma }asin^{2}(\theta )\right) \frac{H}{K\sqrt{F}} \\
\xi \equiv i\left( \frac{J}{\sqrt{K}}+\frac{qer}{\Sigma \sqrt{K}}%
asin^{2}(\theta )\right) .
\end{gather}%
Once again, the second set of Rarita-Schwinger equations (\ref{eqn:RS2b})
will give us additional relations between the various vector components of
the wave function, which are not important here as our solution for the
action will be independent of such relations: 
\end{subequations}
\begin{subequations}
\begin{gather}
\frac{c_{1}}{\sqrt{F}}+\sqrt{g}c_{2}+\frac{d_{3}}{\sqrt{\Sigma }}+\frac{%
Hc_{4}}{K\sqrt{F}}-\frac{id_{4}}{\sqrt{K}}=0 \\
\frac{d_{1}}{\sqrt{F}}-\sqrt{g}d_{2}+\frac{c_{3}}{\sqrt{\Sigma }}+\frac{%
Hd_{4}}{K\sqrt{F}}+\frac{ic_{4}}{\sqrt{K}}=0 \\
\frac{-a_{1}}{\sqrt{F}}+\sqrt{g}a_{2}+\frac{b_{3}}{\sqrt{\Sigma }}-\frac{%
Ha_{4}}{K\sqrt{F}}-\frac{ib_{4}}{\sqrt{K}}=0 \\
\frac{-b_{1}}{\sqrt{F}}-\sqrt{g}b_{2}+\frac{a_{3}}{\sqrt{\Sigma }}-\frac{%
Hb_{4}}{K\sqrt{F}}+\frac{ia_{4}}{\sqrt{K}}=0.
\end{gather}%
Equations (\ref{eqn:BH}) may be rewritten in a way similar to (\ref%
{eqn:fermionaction}), such that they become independent of the wave function
components $a_{\mu },\ldots ,d_{\mu }$ : 
\end{subequations}
\begin{equation}
(\zeta -W_{r}\sqrt{g})(\zeta +W_{r}\sqrt{g})-\left( \frac{W_{\theta }}{\sqrt{%
\Sigma }}-\xi \right) \left( \frac{W_{\theta }}{\sqrt{\Sigma }}+\xi \right)
+m^{2}=0.  \label{eqn:BHsol}
\end{equation}%
Since the action may be reduced to an infinitesimal path from inside to
outside the horizon, $\theta $ may only take values between $\theta
_{0}-\varepsilon $ and $\theta _{0}+\varepsilon $. Hence, we make the
approximation $\theta =\theta _{0}$, thus assuming that $\theta $ is a
constant of motion, which in turns forces $W_{\theta }$ to be constant.
Then, expanding (\ref{eqn:BHsol}) near the horizon $r\rightarrow r_{+}$ and
solving for $W_{r}$, we see that only $\zeta $ contributes, as it rapidly
increases to infinity near the horizon while other terms do not. This allows
us to find a solution similar to that for emission of Dirac particles \cite%
{Fermion tunnelling}: 
\begin{equation}
W_{r\pm }=\frac{\pm 1}{\sqrt{g}}\zeta =\frac{\pm \left( E-J\Omega _{H}+\frac{%
qer_{+}}{r_{+}^{2}+\alpha ^{2}}\right) }{\sqrt{F_{r}(r_{+})g_{r}(r_{+})}%
(r-r_{+})}.  \label{eqn:WrKN}
\end{equation}%
Integrating (\ref{eqn:WrKN}) yields 
\begin{equation}
W_{\pm }=\frac{\pm \pi i\left( E-J\Omega _{H}+\frac{qer_{+}}{%
r_{+}^{2}+\alpha ^{2}}\right) (r_{+}^{2}+\alpha ^{2})}{2r_{+}-2M}.
\end{equation}%
which is independent of the particular choice of \ $\theta _{0}$. Thus,
taking again $\Gamma \propto \exp (-4ImW_{+})$, we find 
\begin{equation}
\Gamma \propto \exp \left( -2\pi \frac{r_{+}^{2}+\alpha ^{2}}{r_{+}-M}\left(
E-J\Omega _{H}+\frac{qer_{+}}{r_{+}^{2}+\alpha ^{2}}\right) \right) ,
\end{equation}%
giving us the expected Hawking temperature for a charged rotating black
hole, 
\begin{equation}
T_{H}=\frac{1}{2\pi }\frac{r_{+}-M}{r_{+}^{2}+\alpha ^{2}}=\frac{1}{2\pi }%
\frac{\sqrt{M^{2}-\alpha ^{2}-e^{2}}}{2M\left( M+\sqrt{M^{2}-\alpha
^{2}-e^{2}}\right) -e^{2}}.
\end{equation}

\section{Conclusions}

We have shown that gravitinos, as described by the Rarita-Schwinger equation
in a curved spacetime, can tunnel out of event horizons. \ The tunnelling
formalism applies equally well to this situation, and we found that
gravitinos attain the familiar Unruh temperature in a Rindler spacetime. We
likewise recover the Hawking temperature for gravitino radiation from a
Kerr-Newman black hole. All four spin components yield the same temperature
to leading order in a WKB approximation. These results are in agreement with
those obtained by applying this method to spin-1/2 fermions \cite{Fermion
tunnelling,kerner-2008}, as well as with results obtained by using the
Newman-Penrose approach on gravitinos \cite{Jing} radiating from a
Reissner-Nordstr\"{o}m black hole.

\section{Acknowledgements}

This work was supported in part by the Natural Sciences and Engineering
Research Council of Canada.

\end{document}